\documentstyle[12pt]{article}
\renewcommand{\topfraction}{1.0}
\renewcommand{\bottomfraction}{1.0}
\renewcommand{\textfraction}{0.0}
\newlength{\dinwidth}
\newlength{\dinmargin}
\newcommand{\resection}[1]{\setcounter{equation}{0}\section{#1}}

\begin{document}

\def\lq{\left [}
\def\rq{\right ]}
\def\LL{{\cal L}}
\def\VV{{\cal V}}
\def\AA{{\cal A}}

\newcommand{\be}{\begin{equation}}
\newcommand{\ee}{\end{equation}}
\newcommand{\bea}{\begin{eqnarray}}
\newcommand{\eea}{\end{eqnarray}}
\newcommand{\nn}{\nonumber}
\newcommand{\dd}{\displaystyle}

\thispagestyle{empty}
\vspace*{4cm}
\begin{center}
  \begin{Large}
  \begin{bf}
TWO BODY NON LEPTONIC DECAYS OF $B$ AND $B_s$ MESONS$^*$\\
  \end{bf}
  \end{Large}
  \vspace{10mm}
  \begin{large}
A. Deandrea, N. Di Bartolomeo and R. Gatto\\
  \end{large}
D\'epartement de Physique Th\'eorique, Univ. de Gen\`eve\\
  \vspace{5mm}
  \begin{large}
G. Nardulli\\
  \end{large}
Dipartimento di Fisica, Univ.
di Bari\\
I.N.F.N., Sezione di Bari\\
  \vspace{5mm}
\end{center}
  \vspace{2cm}
\begin{center}
UGVA-DPT 1993/07-824\\\
hep-ph/9308210\\\
BARI-TH/155\\\
July 1993
\end{center}
\vspace{2cm}
\noindent
$^*$ Partially supported by the Swiss National  Science Foundation
\newpage
\thispagestyle{empty}
\begin{quotation}
\vspace*{5cm}
\begin{center}
  \begin{Large}
  \begin{bf}
  ABSTRACT
  \end{bf}
  \end{Large}
\end{center}
  \vspace{5mm}
\noindent
We perform an analysis of two-body non leptonic decays of $B$ and $B_s$
mesons in the factorization approximation. We make use of the semileptonic
decay amplitudes  previously calculated on the basis of an effective
lagrangian satisfying chiral and heavy quark symmetries and including spin 1
resonances. Exclusive semileptonic $D$-decay data are used as experimental
input. Our results compare favorably with data, whenever they are available
and indicate a positive value for the ratio of non leptonic coefficients
$a_2/a_1$, similarly to previous studies.
\end{quotation}
\newpage
\setcounter{page}{1}

Chiral and heavy quark symmetries \cite{hqet} \cite{georgi} have been used to
construct an effective lagrangian for light and heavy mesons including
their effective weak interactions \cite{chiral}. This approach has been
extended \cite{noi1} to include the nonet of
$1^-$ vector meson resonances, coupled through the so-called hidden gauge
symmetry,  so that one is able to
describe the interactions among heavy $Q{\bar q}_a$ mesons ($Q$ heavy quark,
$a=1,2,3$ for $u,d,s$), the pseudoscalar and the
light vector meson resonances.
In reference \cite{noi}, we performed an extensive analysis of
heavy meson semileptonic decays into final states containing light mesons. The
main result of the analysis was that,
using experimental
information on
$D \to \pi l\nu_l$, $D \to K l\nu_l$ and $D \to K^* l\nu_l$ as  inputs,
together with the hypothesis of a simple pole behaviour of the form factors in
the whole region of $q^2$,
one can make predictions on the form factors
describing the semileptonic decays of the $B$ and $B_s$ mesons into light
mesons.

In the present letter we wish to apply these results to the evaluation of
two body non leptonic decays \cite{bigi}
 of $B$ and $B_s$ within the factorization
approximation. Before going into the details of our work, let us
discuss the meaning and the possible limits of the approximation.

\resection{Non leptonic amplitudes}

Let us first write down the weak non leptonic hamiltonian we shall employ in
our calculations. To be specific let us
 consider $\Delta B =1$, $\Delta S =\Delta
C =0$ transitions (other relevant cases can be handled similarly).
 We take \cite{galee}
\be
H_{NL}=\frac{G_F}{\sqrt{2}}\sum_{q=u,c} V_{qb} V_{qd}^* \left[ C_1 (\mu)O_1^q
+C_2 (\mu)O_2^q \right]
\label{1}
\ee
where the four local operators $O_1^u$, $O_2^u$, $O_1^c$ and $O_2^c$ are
\bea
O_1^u &=& {\bar u}_i \gamma^\mu (1-\gamma_5)b^i {\bar d}_j \gamma_\mu
(1-\gamma_5) u^j
\label{2}  \\
O_2^u &=& {\bar u}_i \gamma^\mu (1-\gamma_5)b^j {\bar d}_j \gamma_\mu
(1-\gamma_5) u^i
\label{3}
\eea
and $O_1^c$ and $O_2^c$  are obtained by the change $u \to c$. In \ref{2}
and \ref{3} $i$ and $j$ are colour indices, whereas in \ref{1} $C_1 (\mu)$ and
$C_2 (\mu)$ are QCD coefficients computed at the scale $\mu \approx m_b$. We
have neglected, in writing eq. \ref{1}, the contribution of the QCD penguin
operators $O_3,\ldots O_6$ since their Wilson coefficients $C_3, \ldots C_6$
are numerically very small as compared to $C_1$ and $C_2$ \cite{buras}; while
these operators could be relevant in some processes where the current-current
operators $O_1$ and $O_2$ are not active or strongly Cabibbo suppressed,
 in the applications we shall consider
here their contribution would never exceed $10 \%$ of the result making
their neglect  reasonable.

The values of the coefficients  $C_1$ and $C_2$ have been computed in
\cite{buras} beyond the leading-log approximation; for $m_b=4.8~GeV$,
$m_t=150~GeV$ and $\Lambda_{{\bar MS}}=250~MeV$ one finds $C_1=1.133$ and
$C_2=-0.291$.

In order to discuss the factorization approximation let us be specific and
consider, e.g., the decay ${\bar B}^0 \to \pi^+\pi^-$. To compute
$<\pi^+\pi^-|H_{NL}|{\bar B}^0>$ we have to consider the two hadronic matrix
elements $<\pi^+\pi^-|O_1|{\bar B}^0>$ and $<\pi^+\pi^-|O_2|{\bar B}^0>$. As
for the first one, the factorization procedure amounts to write ($\Gamma_\mu
=\gamma_\mu (1-\gamma_5)$):
\be
<\pi^+\pi^-|O_1|{\bar B}^0>=<\pi^+|{\bar u}\Gamma_\mu b|{\bar B}^0>
<\pi^-|{\bar d}\Gamma^\mu u|0>
\label{5}
\ee
This approximation can be given
some theoretical justification on the basis of $1/N_c$ expansion
\cite{buras2} or color transparency \cite{bjorken}. Moreover, from a
phenomenological point of view there are indications that
in the realm of B physics, factorization  roughly works
satisfactorily\cite{bortostone}.
As to the second operator,
we can write
\be
<\pi^+\pi^-|O_2|{\bar B}^0>=
<\pi^+\pi^-|\frac{1}{N_c} O_1+\tilde{O}_2|{\bar B}^0>
\label{6}
\ee
with
\be
\tilde{O}_2= {\bar u} \Gamma^\mu T^a b {\bar d} \Gamma_\mu
T^a u
\label{7}
\ee
In \ref{7} $T^a ~(a=1,\ldots 8)$ are colour matrices, normalized according to
$TrT^a T^b=\frac{1}{2} \delta^{ab}$ and the sum over repeated indices is
understood. In the naive factorization, inserting the vacuum, one finds that
the
$\tilde{O}_2$ operator, which contains
 coloured currents, does not contribute
and therefore the contribution of $H_{NL}$ to the amplitude will be given only
by the $O_1$ operator with a multiplicative coefficient $a_1=C_1+\frac{1}{N_c}
C_2$. However, as discussed in \cite{BSW} and \cite{desphande}, the rule of
discarding the operators with coloured currents while applying the vacuum
saturation is ambiguous and unjustified. This is already one reason,
among many others,
to make the choice \cite{BSW} \cite{neubertstech}
to treat $a_1$ and the analogous parameter $a_2$, multiplying $O_2$ in the
naive factorization approximation,  $a_2=C_2+C_1/N_c$, as free parameters.
This will be our attitude in this letter as well.

 Let us recall
that the analysis of $D$ non leptonic decays leads to the empirical finding
$a_1 \approx C_1 , a_2 \approx C_2$ \cite{BSW}.
There has been some recent theoretical effort to understand the empirical rule
of ``discarding the $1/N_c$ term ", which has shown  that, at least
in some channels (e.g. ${\bar B}^0 \to D^+\pi^-$ ) and within a certain
kinematical approximation (i.e. $M_B, M_D \to \infty$ while their difference is
kept fixed) there is a tendency to  cancellation between  the two terms
appearing in \ref{6} \cite{shifman}. Since these analyses
are far from having reached a definite conclusion, we are
justified in our choice, especially because from our analysis, as
we shall discuss below, a positive value of $a_2$ appears favourite,
a result that seems general \cite{stone} and
would indicate that the positive $1/N_c$ term cannot be neglected.

In such a type of calculation one has to be aware of the many uncertainties
and incompleteness heavily bearing down at practically all the levels
of the procedure. Such uncertainties and incompleteness are unfortunately
in a way or in another common to all calculations of such complex
phenomenon as non leptonic decays, and like everybody else we cannot
avoid them.

The standard approach to non leptonic decays consists, as we
have already described, in incorporating short distance effects into an
effective hamiltonian constructed through operator product expansion
and use of renormalization group, while leaving long distance QCD effects
within the hadronic matrix elements. Already the construction of the
effective hamiltonian contains uncertainties, particularly in cases where
many different scales are present, such as for transitions from $b$ to
$c$. One has to add to this the general impossibility of obtaining a scale
independent result when the matrix elements of the scale-dependent
effective hamiltonian are taken within some hadronic model with usually
unrelated scale dependence. One can only hope that, within a class of
processes, one may choose a suitable scale at which this procedure
approximately applies.

The idea behind  the factorization approximation is that hadronization
appears only after the amplitude takes the form of a product of matrix
elements of quark currents which are singlets in color, thus allowing
for approximate deductions from semileptonic processes. Different
kinematical situations may suggest that factorization may apply better
to some non leptonic processes rather than to others.
For instance one intuitively expects that it may work better when a color
singlet current directly produces an energetic  meson easily escaping
interaction with the other quarks. Independent of this, various other effects
such as more or less strong role of long-distance contributions,
including final state interactions effects,
 of small annihilation terms, more or less
sensitivity to choice of the scale, etc., may suggest that the uniform
simultaneous application of the factorization approximation to different
processes must be subject to detailed qualifications and may at the end turn
out to be incorrect. Unfortunately at the present stage of the subject one
is forced to first collect informations by comparing an admittedly very
rough procedure to the data available.

We also note that the effective lagrangian approach can be used to write down
the non leptonic hamiltonian in terms of the heavy and light meson fields
\cite{kamal}, however in this way one can make prediction only for a
small region of the phase space where the light mesons are soft.

According to what already said,
in the factorization approximation, two body non leptonic decays of $B$ and
$B_s$ mesons are obtained by the semileptonic matrix elements of the weak
currents between different mesons. A suitable form is
\be
<P(p')|V^{\mu}|B(p)> =
 \big[ (p+p')^{\mu}+\frac{M_P^2-M_B^2}{q^2} q^{\mu}\big]
F_1(q^2) -\frac {M_P^2-M_B^2}{q^2} q^{\mu} F_0(q^2)
\label{10}
\ee
\bea
<V (\epsilon,p')| &(&V^{\mu}-A^{\mu})|B(p)> =
\frac {2 V(q^2)} {M_B+M_V}
\epsilon^{\mu \nu \alpha \beta}\epsilon^*_{\nu} p_{\alpha} p'_{\beta} \nn\\
&+& i  (M_B+M_V)\left[ \epsilon^*_\mu -\frac{\epsilon^* \cdot q}{q^2}
q_\mu \right] A_1 (q^2) \nn\\
& - & i \frac{\epsilon^* \cdot q}{(M_B+M_V)} \left[ (p+p')_\mu -
\frac{M_B^2-M_V^2}{q^2} q_\mu \right] A_2 (q^2) \nn \\
& + & i \epsilon^* \cdot q \frac{2 M_V}{q^2} q_\mu A_0 (q^2)
\label{11}
\eea
where $P$ is a light pseudoscalar meson and $V$ a light vector meson, $q=p-p'$,
\be
A_0 (0)=\frac {M_V-M_B} {2M_V} A_2(0) + \frac {M_V+M_B}
{2M_V} A_1(0)
\label{12}
\ee
and $F_1 (0)=F_0 (0)$.

For the $q^2$ dependence of all the form factors we have assumed a simple pole
formula $F(q^2)= F(0)/(1-q^2/M_P^2)$ with the pole mass $M_P$ given by the
lowest lying meson with the appropriate  quantum numbers  ($J^P=0^+$ for
$F_0$, $1^-$ for $F_1$ and $V$, $1^+$ for $A_1$ and $A_2$, $0^-$ for $A_0$ ).

The values of the form factors at $q^2=0$, as  given by the study of the
semileptonic decays as performed in \cite{noi},
 are reported in Table I. The errors in the Table follow from the experimental
semileptonic exclusive $D$ decays, that we used as inputs.
In Table
II we report the values of the pole masses we employ. Let us explicitly stress
that the values reported in Table I were obtained
 in the leading order of
the $1/M_Q$ expansion of the heavy chiral current.
 In \cite{noi} we reported also the results of a fit
obtained introducing mass corrections to the leptonic decay constant ratio
$f_B/f_D=\sqrt{M_D/M_B}$ (modulo logs)
predicted by the heavy quark effective theory.
However, as discussed below (see also \cite{noi2}), a few non leptonic B decays
seem to disagree with this latter solution and therefore we choose to
work everywhere at the leading order in $1/M_Q$ expansion.

As for the current matrix elements between a pseudoscalar meson $P$ or vector
meson $V$ we take:
\bea
<0|A^\mu|P(p)> &=& i f_P p^\mu \nn \\
<0|V^\mu|V(p, \epsilon )> &=&  f_V M_V \epsilon^\mu
\label{16}
\eea
In Table III we present the values of $f_P$ and $f_V$ used in computing
the rates. We observe that recent CLEO data on the decay $D_s \to \mu \nu$
point to a rather large value of $f_{D_s}: f_{D_s}= 315 \pm 46$ MeV
\cite{stone}. On the other hand all theoretical approaches indicate
for $f_{D}$ and $f_{D_s}$  smaller values;
for example QCD sum rules \cite{dominguez}
give $f_{D} = 223 \pm 27 MeV$ and
$f_{D_s} = 277 \pm 13 MeV$; a recent QCD sum rules analysis
for the ratio $f_{D_s}/f_{D}$ gives the value $\approx 1.1$ \cite{blasi}
and
similar results
are obtained from Lattice QCD or potential models (for a review
see \cite{nardulli}).
In view of these results we have assumed for $f_{D}$ and $f_{D_s}$
the values reported in Table III  that are compatible with both CLEO data
and theoretical expectations.

\resection{Numerical results}

Let us begin
 our numerical analysis by discussing a class of decays that depend
only on the parameter $a_2$. Recent data from CLEO Collaboration \cite{stone}
allow for  a determination of this parameter. From
$BR(B \to K J/\psi)=(0.10 \pm 0.016) \times 10^{-2}$,
$BR(B \to K^* J/\psi)=(0.19 \pm0.036) \times 10^{-2}$ and
$BR(B \to K \psi(2s))=(0.10 \pm 0.036) \times 10^{-2}$ we obtain
\be
|a_2| \simeq 0.27
\ee
Using this value we can compute the results of Table IVa that contain
two body non leptonic ${\bar B}^0$ decays  depending only on
$a_2$ and on current matrix elements between $B$ and light mesons.
We have used the recent determination for $V_{ub}$ arising from
the CLEO analysis of the  end-point of lepton spectrum in semileptonic
inclusive $B$ decays \cite{soling}: $V_{ub}/V_{cb}=0.075$. We take
$V_{cb}=0.040$.

Let us now study the coefficient $a_1$. The best way to determine it
is to consider
${\bar B}^0$ decays into $D^{*+} \pi^-$, $D^{*+} \rho^-$,
$D^{+} \pi^-$, $D^{+} \rho^-$ final states. In order to use the experimental
data we need an input for current matrix elements between
$B$ and $D, D^*$ states.

In \cite{noi} we did not
 consider $B \to D$ and $B \to D^*$ transitions in the heavy
quark effective theory; this subject has been recently actively investigated
by several authors and we rely on their work to compute the
corresponding non leptonic decay rates.
The relevant matrix elements at leading order in
$1/M_Q$ are \cite{georgi}
\be
<D(v')|V^{\mu}(0)|B(v)> =
\sqrt{M_B M_D} \xi (w) [v+v']^\mu
\label{13}
\ee
\bea
& & <D^* (\epsilon,v')| (V^{\mu}-A^{\mu})|B(v)>  = \nn \\
& &\sqrt{M_B M_D} \xi (w)
\left[ - \epsilon^{\mu\lambda\rho\tau} \epsilon^*_\lambda v_\rho v'_\tau +
i(1+ v \cdot v')\epsilon^{*\mu} -i(\epsilon^* \cdot v)v'^{\mu} \right]
\label{14}
\eea
where $w=v \cdot v'=(M_B^2+M_D^2-q^2)/2 M_B M_D$, $v$ and $v'$ are the
meson velocities, and $\xi (w)$ is the Isgur-Wise form factor. $\xi (w)$ has
been computed by QCD sum rules \cite{sumrul} \cite{sumrul1}, potential models,
and estimated phenomenologically \cite{neubertstech}.  We shall take
for it the expression \cite{neubert3}
\be
\xi (w) =\left( \frac{2}{1+w} \right) \exp \left[
- (2 \rho^2-1)\frac{w-1}{w+1}\right]
\label{14bis}
\ee
which reproduces rather well
the semileptonic data\cite{data} with $\rho \simeq 1.19, V_{cb}=0.04,
\tau_B=1.4 ps$. We stress
once again that we have chosen to work at the leading order in $1/M_Q$, which
is why we have not introduced the non leading form factors discussed e.g. in
\cite{luke} \cite{neubert}.

{}From the new CLEO data \cite{stone}
$BR({\bar B}^0 \to D^+ \pi^-) = (2.2 \pm 0.5 )\times 10^{-3}$,
$BR({\bar B}^0 \to D^{*+} \pi^-) = (2.7 \pm 0.6 )\times 10^{-3}$ ,
$BR({\bar B}^0 \to D^+ \rho^-) = (6.2  \pm 1.4 )\times 10^{-3}$ and
$BR({\bar B}^0 \to D^{*+} \rho^-) = (7.4  \pm 1.8 )\times 10^{-3}$
 one gets
\be
|a_1| \simeq 1.0
\ee
For completeness we have reported
in Table IVb the results for non leptonic decays depending only on $a_1$
and on the Isgur-Wise form factor $\xi(w)$.

In Table IVc we use the fitted value of $a_1$ together with
the weak matrix elements computed in our model to compute additional two body
decays depending on $a_1$. We observe explicitly that the recent
CLEO result \cite{kim} $BR(B^0 \to \pi^+ \pi^-)= (1.3^{+0.8}_{-0.6}\pm 0.2)
\times 10^{-5}$
is in agreement with our outcome. We stress that this agreement depends
on our choice of the "scaling" solution for the semileptonic matrix
elements, according to our previous discussion. The other solution discussed
in ref.\cite{noi} appears to be excluded by the present data for this
decay channel.

In Table V we consider $B^-$ decays.
Let us observe that
they depend on the relative sign between $a_1$ and $a_2$
which has not been fixed yet. The new CLEO data \cite{stone}
$BR(B^- \to D^0 \pi^-) = (4.7 \pm 0.6 )\times 10^{-3}$,
$BR(B^- \to D^{*0} \pi^-) = (5.0 \pm 1.0 )\times 10^{-3}$ ,
$BR(B^- \to D^0 \rho^-) = (10.7 \pm 1.9 )\times 10^{-3}$ and
$BR(B^- \to D^{*0} \rho^-) = (14.1 \pm 3.7 )\times 10^{-3}$,
allow to conclude that the ratio $a_2/a_1$ is
positive.
Clearly this result
depends on the relative phase of the hadronic matrix elements. Our
choice is the ``natural" one, i.e. we assume (analogously to
\cite{BSW}) that
for a decay $B \to M_1 M_2$ ($M_1$ and $M_2$ scalar or vector mesons)
the phase is the one determined under
the assumption of spin and flavour symmetry in
the meson spectrum. Of course these symmetries are (even
badly) broken in many decays; one should therefore be aware of the
possibility to have a different phase between the two terms in such cases.

Our results for $B^-$ decays are reported in Table V, while
in Table VI we compute analogous results for the $B_s$ non leptonic
decay channels.
In doing the calculations we have neglected
the mixing $\eta$-$\eta'$, and we have assumed ideal mixing between $\omega$
and $\phi$. Moreover we have
not taken into account final state interactions, whose effect
we cannot evaluate at the moment (in $D$ decays, as shown in \cite{BSW},
their effect was in some cases rather significant), and all effects of mixing
and CP violation.

We also note that our reported theoretical errors in Tables IV, V and VI
only reflect the uncertainties in the weak couplings reported in Table I.
Since the form factor $A_0$ at $q^2=0$
has an uncertainty of about $100\%$ (see Table I), the
$BR$'s containing this coupling have a large
error, as one can see in some entries of the Tables IV-VI
and should therefore  be taken with care.

 Let us finally comment on the decays with two charmed mesons in the
final state (see Tables IVb, Va and VIb). Our results are in agreement
with data from both CLEO and ARGUS collaboration \cite{pdg} \cite{stone}.
According to the discussion of ref. \cite{grinstein}, however, the correct
value of the $a_1$ coefficient to be used in connection with the
the factorization hypothesis and the
Heavy Quark Effective Theory is $|a_1| \simeq 1.45$. Indeed, as discussed
in \cite{grinstein}, since the light quarks do not carry large momenta in these
decays, the running strong coupling constant should be computed at a
scale lower than $m_c$, which results in a value for the $|a_1|$ coefficient
significantly larger than the one used in Tables IV-VI. Using experimental data
on decays into $D^{(*)}D_s^{(*)}$ \cite{pdg} \cite{stone}, for
example ($BR({\bar B}^0 \to D^+ D^-_s)= (0.6 \pm 0.45)\times 10^{-2}$)
one can see that the above mentioned
large value of $|a_1|$ is incompatible with the data. It
would be interesting to analyze if such disagreement reflects
a breakdown of the factorization approximation or the presence of
non leading effects in $1/m_Q$ approximation.
\par
In conclusion, we have performed, in the factorization approximation,
an analysis of two body non leptonic decays
of the $B$ and $B_s$ mesons; our study has been based on
semileptonic amplitudes obtained by an effective lagrangian
having chiral and heavy quark symmetries and has employed semileptonic
exclusive $D$ decays as an input. Our results are in agreement with
the experimental data, whenever they are available, and indicate,
similarly to other analyses, a positive value for the ratio
of the non leptonic coefficients $a_2/a_1$. Our results represent,
in our opinion, a preliminary indication that
 B semileptonic decays to light mesons can be related to the
analogous D decays without major violations
of the heavy quark flavour symmetry.
\par
\vspace*{1cm}
\noindent
{\bf  Acknowledgement}:
We would like to thank Professor S. Stone for discussions on the CLEO
results.
We would like to thank R. Casalbuoni, P. Colangelo and F. Feruglio
for frequent discussions and exchanges during the course of this work.

\newpage
\begin{center}
  \begin{Large}
  \begin{bf}
  Tables Captions
  \end{bf}
  \end{Large}
\end{center}
  \vspace{5mm}
\begin{description}

\item[ Table I] Form factors at zero momentum transfer for $B \to P$ and $B \to
V$ semileptonic transitions (see ref.\cite{noi}).
Indicated errors are from the experimental inputs.
 The form factors $A_1$ and $A_2$ are relatively well known, while
$A_0$ has a relative error  $\sim 100\%$.

\item [Table II] Pole masses for different states. Units are $GeV$.

\item [Table III] Values of leptonic decay constants. For particles that can
couple to different weak currents, the corresponding weak current is indicated
within brackets. For $\phi$ and $\omega$ we assume ideal mixing, while
$\eta$ is the pure octet component.

\item[Table IV] Predicted widths and branching ratios for ${\bar B}^0$ decays.
We use in the Tables  $\tau_{{\bar B}^0}=$ $\tau_{B_s}=$
$\tau_{B^-}=14 \times 10^{-13} s$, $V_{ub}=0.003$, $V_{cb}=0.04$.
The quoted errors come only from the uncertainties of the form factors of
Table I

\item[Table V] Predicted widths and branching ratios for $B^-$ decays.

\item[Table VI] Predicted widths and branching
 ratios for ${\bar B}^0_s$ decays.

\end{description}

\newpage
\begin{table}
\begin{center}
\begin{tabular}{l c c c c c }
\hline
Table I & & & & & \\
\hline
Decay & $F_1=F_0$ &  $V$ & $A_1$ & $A_2$ & $A_3=A_0$
 \\ \hline
$B \to K$ & $0.49\pm 0.12$ & & & & \\
$B \to \pi^{\pm}$ & $0.53\pm 0.12$ & & & & \\
$B \to \eta$ & $0.49 \pm 0.12$ & & & & \\
$B_s \to \eta$ & $0.52 \pm 0.12$& & & & \\
$B_s \to K$ & $0.52 \pm 0.12$& & & & \\
$B \to \rho^{\pm}$ &  &$0.62 \pm 0.12$ &$0.21\pm 0.02$ &$0.20\pm 0.08$ &$0.24
 \pm 0.24$ \\
$B \to \omega$ &  &$0.62 \pm 0.12$ &$0.21 \pm 0.02$ &$0.20\pm 0.08$ &
 $0.24 \pm 0.24$ \\
$B \to K^*$ &  &$0.61 \pm 0.12$ &$0.20 \pm 0.02$ &$0.20 \pm 0.08$ &
 $0.20 \pm 0.20$ \\
$B_s \to K^*$ &  &$0.64\pm 0.12$ &$0.20\pm 0.02$ &$0.21\pm 0.08$ &
$0.17 \pm 0.21$ \\
$B_s \to \phi$ &  &$0.62 \pm 0.12$ &$0.19\pm 0.02$ &$0.20\pm 0.08$ &
$0.17 \pm 0.18$ \\
\hline \hline
\end{tabular}
\end{center}
\end{table}

\vspace{2cm}
\begin{table}
\begin{center}
\begin{tabular}{l c c c c}
\hline
Table II & & & & \\
\hline
State $J^P$ & $0^-$ & $1^-$ & $0^+$ & $1^+$ \\ \hline
${\bar d} c$ & 1.87 & 2.01 & 2.47 & 2.42 \\ \hline
$ {\bar s} c$ & 1.97 & 2.11 & 2.60 & 2.53 \\ \hline
$ {\bar u} b$ & 5.27 & 5.32 & 5.78 & 5.71 \\ \hline
$ {\bar s} b$ & 5.38 & 5.43 & 5.89 & 5.82 \\
\hline \hline
\end{tabular}
\end{center}
\end{table}

\vspace{50 mm}
\begin{table}
\begin{center}
\begin{tabular}{l c l c }
\hline
Table III & & & \\
\hline
Particle & $f_P$ & Particle & $f_V$ \\ \hline
$\pi^{\pm}$ & 132 & $\rho^{\pm}$ & 221 \\
$K$ & 162 & $K^*$ & 221 \\
$D$ & 240 & $ D^*$ &  240 \\
$D_s$ & 270 & $D^*_s$ & 270 \\
$\eta (u{\bar u} / d{\bar d})$ & 54 & $J/\psi$ & 409 \\
$\eta (s{\bar s})$ & -108 & $\phi$ & 221 \\
 & & $\omega$ & 156 \\
\hline \hline
\end{tabular}
\end{center}
\end{table}
\end{document}